\documentclass[english,aps]{revtex4}
\usepackage{babel}
\usepackage{graphicx}
\usepackage{amsmath}
\begin{document}






\title{Operator Method for Nonperturbative Calculation of the Thermodynamic
Values in Quantum Statistics. Diatomic Molecular Gas.}
\author{I.D.Feranchuk$^1$\footnote[1] {fer@open.by} and A.A.Ivanov$^{1,2}$}

\address{ $^1$
Department of Theoretical Physics, Belarusian State University, \\
4 Fr.Skariny av., 220080 Minsk, Republic of Belarus \\ \\
$^2$ Department of Natural Sciences, Belarusian National Technical University, 65 Fr.Skariny av., 220013 Minsk, Republic of Belarus}

\begin{abstract}
\noindent \textbf{Abstract}\\Operator method and cumulant expansion are used for nonperturbative calculation of
the partition function and the free energy in quantum statistics. It is shown for Boltzmann diatomic
molecular gas with some model intermolecular potentials that the zeroth order approximation of the proposed method
interpolates the thermodynamic values with rather good accuracy in the entire range of both the Hamiltonian
parameters and temperature. The systematic procedure for calculation of the corrections to the zeroth order
approximation is also considered.
\end{abstract}

\maketitle
\newpage


Short Title: Operator Method in Quantum Statistics


\maketitle

\newpage

\section{Introduction}

At present, most physical problems of interest can be solved neither exactly nor in terms of the canonical
perturbation theory (CPT), therefore a considerable attention is given to the development of methods of
nonperturbative analysis of quantum systems (see, e.g., review \cite{1a} and references therein). One of such
methods, called the operator method (OM) of approximate solution of Schr\"{o}dinger equation, proved to be very
effective. The OM was presented in \cite{1b} and approved for a number of physical problems including those
with many degrees of freedom \cite{2a} - \cite{2d}. The main advantages of the OM are defined by the fact that its
zeroth order approximation allows one to calculate the Hamiltonian eigenvalues and eigenvectors with rather high
accuracy in the entire range of both the Hamiltonian parameters and quantum numbers of the states. Besides, the
further approximations of the OM lead to the sequences converging to the exact solutions. However, most of the
particular applications of the OM have been considered before for the systems in \textquotedblleft
pure\textquotedblright\ quantum states.

The nonperturbative methods are of great interest also in quantum statistics, when calculating thermodynamic
characteristics of quantum systems. These characteristics can be expressed through either the partition function
or the free energy of a system. Most of known nonperturbative methods in quantum statistics are based Feynman path integrals (e.g., \cite{2e} and
references therein). However, an alternative representation of the partition function based on the direct
summation over the energy states is also important because it could be more convenient and descriptive for some
applications, especially in the atomic and molecular physics when the system is characterized by a large but
finite number of degrees of freedom. For example, the additional procedure of bosonization is necessary when the
functional integrals are used for the spin systems, and that leads to complication of the real Hamiltonian (see,
e.g., \cite{2g}).

The purpose of this paper is to generalize the OM for the nonperturbative calculation of the thermodynamic
values in quantum statistics with the Schr\"odinger representation for quantum systems. A specific feature of
this problem in comparison with application of the OM for ``pure'' quantum states is that the partition function
is defined by an energy spectrum as a whole and include the summation over all states. As a result, the
accuracy of calculation of the thermodynamic values for some temperature interval can be rather low even in the
case when the Hamiltonian eigenvalues are found with high precision (see example in Sec.2). Therefore we
should analyze whether the OM accuracy for the spectral problem \cite{1} ensures the calculation of the
thermodynamic values in the entire temperature range. Besides, some additional procedure should be developed for
approximate summation over the states. It is shown in the paper that the regular nonperturbative method for
calculation of the thermodynamic values can be developed on the basis of the OM combined with the cumulant
expansion (CE) \cite{2}. In order to approve our approach and compare it with other approximate methods, the
Boltzmann diatomic molecular gas is considered and the partition function and the free energy corresponding to the
molecular internal degrees of freedom are calculated. It is shown for some particular examples that the zeroth
order approximation of the proposed method leads to the uniformly suitable estimation for the thermodynamic
values in the entire range of both temperature and Hamiltonian parameters. The systematic procedure for
calculation of the consequent approximations is also considered.

The paper is organized as follows: in Sec. 2 we discuss some general definitions of uniformly suitable
estimation that concerns any nonperturbative approach. The procedure for nonperturbative calculation of the
Hamiltonian eigenvalues on the basis of the operator method is described in Sec.3 as the first part of the
proposed approach. The approximate summation over the states by means of the cumulant expansion is considered in
Sec.4 as the second part of our method. In order to approve this expansion, the analytical formula for
interpolation of the rotational partition function of the diatomic molecular gas is obtained in the same section
on the basis of the method proposed. In Sec.5 we use our approach for calculation of the partition function and
the free energy for an equilibrium system of the anharmonic oscillators and compare the calculated values with
the results of other methods. The obtained results are of interest not only as the approval of the considered
method, but also for the description of thermodynamic characteristics of real molecular systems when anharmonic
effects should be taken into account.

\section{Formulation of the Problem}

It is well known that if the Hamiltonian of some quantum system includes a
small parameter, the regular procedure of calculation of the observable
characteristics of this system can always be constructed in the form of a
power series in this parameter. For example, it can be the canonical form of
the perturbation theory with expansion in a power of small coupling constant
($g\ll 1$); an expression as powers in $g^{-1}$ for the strong coupling
limit; a series in terms of powers of $\hbar $ in the quasi-classical
approximation; low- or high-temperature expansions for thermodynamic values.
However, in most cases these expansions are the asymptotic ones, and hence
cannot be used directly in a wide range of the Hamiltonian parameters. On
the contrary, the main goal of the nonperturbative methods is to calculate
physical characteristics of a system in the entire range of its parameters.
Further in the paper, when we refer  to the nonperturbative calculation of
some physical values, we assume that either there is no small parameter in
the system, or its value lies out of the domain of applicability of the
asymptotic expansions \cite{1a} - \cite{1}.

It is convenient to define the concept of the ``uniformly suitable
estimation'' (USE) in this case. Let us consider some physical variable with
eigenvalues $F_n(\lambda)$ depending on the quantum number $n$ and the
physical parameter $\lambda$ (the totality of the parameters and quantum
numbers are implied for systems with many degrees of freedom). Let us
introduce

\noindent \textbf{Definition 1}: the function $F_{n}^{(0)}(\lambda )$ is the
USE for $F_{n}(\lambda )$ if the following inequality holds in the entire
range of variation of the values $n$ and $\lambda $

\begin{equation}  \label{1}
\left\vert \frac{F^{(0)}_n(\lambda) - F_n(\lambda)}{F_n(\lambda)}
\right\vert \leq \xi^{(0)}.
\end{equation}

\noindent Here the parameter $\xi^{(0)} < 1$ is assumed to be independent of
$n$ and $\lambda$, and defines the accuracy of the USE. We also consider

\noindent \textbf{Definition 2}: there are the sequence of functions $%
F_{n}^{(s)}(\lambda )$ and the method for their calculation corresponding to
the decreasing sequence of the parameters $\xi ^{(s)};\quad s=0,1,2,\ldots $%
, so that

\begin{equation}  \label{2}
\lim_{s \to \infty} F^{(s)}_n(\lambda) = F_n(\lambda).
\end{equation}

\noindent It seems that Definition 1 is not constructive because the exact
values $F_{n}(\lambda )$ are unknown in general case. However, there are
several possibilities to estimate the value $\xi ^{(0)}$. In particular, one
can compare the asymptotic series for $F_{n}(\lambda )$ in various limit
cases either of the parameter $\lambda $ or of the quantum number $n$ with
the corresponding expansions of the function $F_{n}^{(0)}(\lambda )$ taking
into account that inequality (\ref{1}) should hold for all cases
simultaneously. Besides, an estimation for $\xi ^{(0)}$ can be found as the
difference between $F_{n}^{(1)}(\lambda )$ and $F_{n}^{(0)}(\lambda )$. For
example, we can refer to the USE for the eigenvalues of the Hamiltonian
calculated for various physical systems in \textquotedblleft
pure\textquotedblright\ states on the basis of the OM \cite{1} - \cite{2f}.

Considerable advantage of the USE for various applications in comparison to
the asymptotic expansions is the possibility to investigate the qualitative
peculiarities of the quantum system in the intermediate range of the
parameters connected with Hamiltonian and the external conditions. At the
same time, rather high precision of the USE in the zeroth order is of great
use for practical calculations and it defines the rate of convergence of the
consecutive approximations to the exact solution.

Let us consider from this point of view the thermodynamic perturbation
theory in the Schr\"odinger representation of the quantum statistics.
Usually it is formulated for the free energy of the system \cite{5} and the
leading terms are the following

\begin{eqnarray} \label{3}
F(\mu ,\beta ) &=&F_{0}+\mu \sum_{n}V_{nn}w_{n}+\mu
^{2}\sum_{n}\sum_{m\not=n}\frac{|V_{mn}|^{2}w_{n}}{E_{n}^{(0)}-E_{m}^{(0)}}
\nonumber \\
&&+\frac{1}{2}\beta \mu ^{2}\left[ \left( \sum_{n}V_{nn}w_{n}\right)
^{2}-\sum_{n}V_{nn}^{2}w_{n}\right] +\cdots .
\end{eqnarray}

\noindent Here we have introduced $\beta =1/kT$ where $T$ is the temperature
and $k$ is the Boltzmann constant; $F$ and $F_{0}$ are the exact and
approximate free energies respectively; $E_{n}^{(0)}$ are the eigenvalues of
the unperturbed Hamiltonian and $V_{mn}$ are the matrix elements of the
perturbation operator with the following form of the total Hamiltonian

\[
\hat H = \hat H_0 + \mu \hat V,
\]

\noindent and $w_{n}=\exp {[}\beta (F_{0}-E_{n}^{(0)}){]}$ is the
unperturbed density matrix.

Any partial sum of this series does not yield the free energy in the whole
range of the temperature and the perturbation parameter $\mu $ even in
the simplest cases. Let us illustrate this by means of the model Hamiltonian
used earlier in \cite{6} for the convergence analysis of a  usual
perturbation series for \textquotedblleft pure\textquotedblright\ states

\begin{eqnarray}  \label{5}
\hat H = \hat H_0 + \mu \hat V = \frac{1}{2}(\hat p^2 + \hat x^2) + \mu \hat x^2; \quad E_n^{(0)} = n
+ \frac{1}{2}; \quad F_0 = \frac{1}{\beta}\ln{[2\sinh{(\beta/2)}]};
\nonumber \\
V_{mn} = (n + \frac{1}{2})\delta_{m,n} + \frac{1}{2}[\sqrt{(n+1)(n+2)}
\delta_{m,n+2} + \sqrt{n(n-1)} \delta_{m,n-2}] ,
\end{eqnarray}

\noindent where $\delta_{m,n}$ is the Kroneker symbol.

Certainly, the exact free energy is well-known for this model ($F=\frac{1}{%
\beta }\ln {[2\sinh {(\beta /2\sqrt{1+2\mu })}]}$) but if we use these
matrix elements in formula (\ref{3}), then rather simple calculation leads
to the following result

\begin{eqnarray}  \label{6}
F(\mu,\beta) = \frac{1}{\beta}\ln{[2\sinh{(\beta/2)}]} + \frac{\mu}{2}%
\coth{\frac{\beta}{2}} -  \nonumber \\
\frac{\mu^2}{8}[1 + \coth{\frac{\beta}{2}} + \frac{1}{2\sinh^2{\frac{\beta}{2%
}}}(1 + 2\beta) + \cdots].
\end{eqnarray}

It is evident that this series does not satisfy the USE criteria with
respect to both essential parameters of the system. In the low temperature
limit ($\beta \rightarrow \infty $), formula (\ref{6}) leads to the power
series of $\mu $ for the ground state energy and this series diverges in the
range of $\mu >1/2$ because of a singular point of the exact eigenvalue $%
E_{n}=\sqrt{1+2\mu }(n+\frac{1}{2})$ in the complex plane of $\mu $ \cite{6}%
. On the other hand, Definition 1 breaks also for the free energy dependence
on the parameter $\beta $. When the temperature increases ($\beta
\rightarrow 0$), the second order correction has a stronger singularity ($%
\sim -\mu ^{2}/4\beta ^{2}$) than the exact free energy.

Thus, our objective is to formulate another regular method, different from (%
\ref{3}), which allows us finding the USE for the free energy of a quantum
system with an arbitrary energy spectrum $E_{n}$.

\section{The Operator Method for the Uniformly Suitable Estimation of the
Eigenvalues}

As mentioned above, in order to obtain the USE for the partition function one has to solve two problems: 1)
to find such an approximate representation for the energy levels of a system under consideration, which
holds in the whole range of the Hamiltonian parameters, and 2) to perform approximate summation on quantum
numbers so that the result remains uniformly suitable with any temperature. For the former purpose we shall
use the operator method (OM) of approximate solution of Schr\"{o}dinger equation. This method is
described in details in review paper \cite{1}. Let us consider here only the basic expressions defining
the consecutive approximations of the OM for the eigenvalues $E_{n}$ and eigenvectors $\left|\psi _{n}\right\rangle$ of a
Hamiltonian $\hat{H}$ of some quantum system  (the index $n$ may involve the whole set of quantum state
numbers)

\begin{eqnarray}  \label{61}
\hat H \left| \psi_n \right\rangle = E_n \left| \psi_n \right\rangle.
\end{eqnarray}

Let us introduce the complete set of state vectors $\left|n,\omega \right\rangle$, depending
on the same quantum numbers and arbitrary parameters $\omega $. In the
canonical perturbation theory (CPT) the eigenfunctions of some part $\hat{H}%
_{0}$ of the full Hamiltonian (zeroth order approximation Hamiltonian) are used as
such a complete set. The Schr\"{o}dinger equation for that part is assumed
to be exactly solvable. In terms of the OM, the vectors $\left|n,\omega \right\rangle$ are
considered as a set of variational wave functions. Thus, the choice of these
vectors is rather arbitrary and is defined by qualitative features of the
system under consideration and by the possibility of rather simple
calculation of the matrix elements of the total Hamiltonian $\hat{H}$. Then
we can represent the solution of equation (\ref{61}) in the form of
expansion

\begin{eqnarray}  \label{62}
\left| \psi_n \right\rangle = \left|n, \omega \right\rangle + \sum_{k \not= n} C_{nk} \left|k, \omega \right\rangle,
\end{eqnarray}

\noindent and choose the normalization of the exact solution as the following

\begin{eqnarray}  \label{63}
\left\langle  \psi_n \right|\left.n, \omega \right\rangle = 1.
\end{eqnarray}

As a result, the initial eigenvalues problem (\ref{61}) exactly reduces to
the system of infinite number of nonlinear algebraic equations for $E_{n}$
and coefficients $C_{nk}$ \cite{1}

\begin{eqnarray}  \label{64}
E_n = H_{nn} + \sum_{k \not= n} C_{nk} H_{nk}; \quad H_{nk} (\omega) = \left\langle n,
\omega \right| \hat H \left|k, \omega \right\rangle;  \nonumber \\
C_{nm} = - [H_{mm} - E_n]^{-1} \bigl[H_{mn} + \sum_{k \not= m,n} C_{nk} H_{mk}\bigr]; \quad m \not= n.
\end{eqnarray}

As shown in \cite{1}, the most effective method for solving this system of
equations is an iterative scheme, in which the exact solutions are
calculated as the limit of the sequences (and not by summation of series, as
in the CPT)

\begin{eqnarray}  \label{65}
E_n = \lim_{s \to \infty} E_n^{(s)}; \quad C_{nk} = \lim_{s \to \infty}
C_{nk}^{(s)},
\end{eqnarray}

\noindent and the consecutive approximations in (\ref{65}) are calculated by
means of recurrence relations

\begin{eqnarray}  \label{66}
E^{(s)}_n(\omega) = H_{nn} + \sum_{k \not= n} C^{(s-1)}_{nk} H_{nk};
\nonumber \\
C^{(s)}_{nm} = - [H_{mm} - E^{(s-1)}_n]^{-1} \bigl[H_{mn} + \sum_{k \not= m,n}
C^{(s-1)}_{nk} H_{mk}\bigr];  \nonumber \\
E^{(0)}_n = E^{(1)}_n = H_{nn}; \quad C^{(0)}_{nk} = 0.
\end{eqnarray}

As shown for various physical systems \cite{1} - \cite{2f}, recurrence
relations (\ref{66}) rapidly converge to the exact solutions, and the choice
of the parameter $\omega $ influences only the rate of this convergence. In
this paper we restrict ourselves with expressions for the energy after the
second iteration (the first one coincides with the zeroth order approximation by
definition), as we are interested mainly not in strict numerical
calculations, but in constructing of the USE for the eigenvalues, the
partition function and the free energy. The second iteration leads to the
following approximation

\begin{eqnarray}  \label{67}
E^{(2)}_n(\omega)\simeq E_n^{(0)}(\omega) + \Delta E_n(\omega) = H_{nn} -
\sum_{m \neq n} [H_{mm} - H_{nn}]^{-1} H_{nm}H_{mn}.
\end{eqnarray}

Strictly speaking, this expression is suitable if the diagonal matrix elements

$$
H_{kk}(\omega) \not= H_{nn}(\omega),
$$
for the considered value of the parameter $\omega$ and for all states from the set $\left|k,\omega\right\rangle$ in
expansion (\ref{62}).

If for some concrete values m and $\omega_m$

$$
H_{mm}(\omega_m) = H_{nn}(\omega_m),
$$
one should extract the state $\left|m,\omega_m\right\rangle$ from expansion (\ref{62}) and solve the secular equation in
the OM zeroth order approximation analogously to the canonical perturbation theory in case of degenerate states
\cite{Landau1}. Some applications of the OM for physical systems with such features were considered in papers \cite{2a}, \cite{2b}.

For systems with several degrees of freedom  the state vector $\left| \psi_n, j \right\rangle$ in equation (\ref{61})
may also depend on some additional index $j$. The energy level $E_n$ is supposed to be independent of this index. Actually it
means that the wave function is also the eigenvector of some operator

\begin{eqnarray}  \label{67a}
\hat J \left| \psi_n, j \right\rangle = j \left| \psi_n, j \right\rangle, \quad \bigl[\hat J, \hat H\bigr] = 0,
\end{eqnarray}

\noindent which commutes with the Hamiltonian. As shown in paper \cite{1}, in such a case the projection operator
$\hat T_j$ for the state with fixed quantum number $j$ should be used in series (\ref{62}) in order to
solve both equations (\ref{61}) and (\ref{67a}) simultaneously. It leads to unessential modification of
recurrence equations (\ref{66}) but does not change the properties of the OM consecutive approximations.

In this paper the generalization of the OM for quantum statistics is of main interest and thus we shall not
consider the above mentioned more complicated systems.

In spite of the formal similarity of formula (\ref{67}) and the expression
obtained in the second order of the CPT (see formula (\ref{3})), the former
one maintains some essential differences. First, the matrix elements of the
operator $\hat{H}$ are calculated with the set of state vectors $\left\vert
n,\omega \right\rangle $ for the whole Hamiltonian (and not for its
unperturbed part as in the CPT). According to \cite{1}, this circumstance is
essential for the convergence of the consequent approximations of the OM.
In particular, this convergence depends on the numerical dimensionless
parameter, defined by the ratio of the diagonal and nondiagonal matrix
elements of the total Hamiltonian

\begin{eqnarray}  \label{68}
\xi (\omega) \simeq \left| \frac{H_{nk}}{H_{nn}} \right|_{max}; \quad k
\not= n.
\end{eqnarray}

This value is independent of the physical parameters of the Hamiltonian and
defines both the convergence of the consequent approximations of the OM, and
the accuracy of the USE for the zeroth order approximation of the OM \cite{1} in
accordance with Definition 1.

Another specific feature of approximation (\ref{67}) is that it depends
on some undefined parameter $\omega $, which should be chosen so that to
provide the best accuracy of the USE in the zeroth order approximation of the OM.
It was shown in \cite{1}, that the best zeroth order approximation of the OM for
the energy level with the quantum number $n$

\begin{eqnarray}  \label{691}
E_n \simeq E_n^{(0)} = E_n^{(0)}(\omega_n),
\end{eqnarray}

\noindent is achieved when $\omega = \omega_n$ is chosen as the solution of
the equation

\begin{eqnarray}  \label{69}
\frac{\partial E_n^{(0)}(\omega)}{\partial \omega} = 0.
\end{eqnarray}

It should be underlined, that equation (\ref{69}) is not the variational
principle for the excited states. The point is that the artificial parameter
$\omega $ defines actually the representation for the wave functions, and
hence, the exact eigenvalues of the Hermitian Hamiltonian should not depend
on its choice. The exact condition is

\begin{eqnarray}  \label{70}
\frac{\partial E_n(\omega)}{\partial \omega} = 0.
\end{eqnarray}

Therefore, equation (\ref{69}) can be considered as the OM zeroth order approximation for the exact condition (\ref{70}).

It is essential that the optimal value of the parameter $\omega _{n}$
depends on the quantum number of the considered state. This means that the
orthonormal set of basic vectors used in expansion (\ref{62}) should be
chosen for every state in different ways. As a result, the consecutive
approximations of the OM (\ref{66}) have \textquotedblleft
local\textquotedblright\ character in the space of quantum numbers, i.e.,
they are calculated independently with different sets of state vectors for
every state of the system. On the contrary, in terms of the CPT the
corrections for all quantum numbers are defined by the same spectrum and
basic state vectors of the unperturbed Hamiltonian.

In this paper we use model physical systems in order to illustrate the
applicability of the OM in statistical physics. In particular, let us
consider the quantum anharmonic oscillator (QAO) with the Hamiltonian

\begin{eqnarray}  \label{71}
\hat H = \frac{1}{2} (\hat p^2 + \hat x^2) + \mu \hat x^2 + \lambda \hat x^4.
\end{eqnarray}

It is well known, that the CPT expansion for this system has zero
radius of convergence with respect to the parameter $\lambda $ \cite{Hioe}. Therefore, the QAO is widely used
for approbation of various nonperturbative methods, and all approximations can be compared with the detailed
numerical analysis of this problem in \cite{Hioe}. In the problem of the QAO the most adequate choice of the
full set of state vectors for the OM algorithm is the set of eigenfunctions of the harmonic oscillator with an
arbitrary frequency, playing a role of the parameter $\omega $. At the same time, it is convenient to perform
all the necessary calculations of the matrix elements in an algebraic form, using the representation of the
secondary quantization \cite{1b}. In such a case, the variational parameter $\omega $ can be introduced
directly into the Hamiltonian by means of creation $\hat{a}^{+}$ and annihilation $\hat{a}$ operators

\begin{eqnarray}  \label{72}
\hat{x} = \frac{1}{\sqrt{2\omega}} (\hat a + \hat a^{+}), \qquad \hat p = -i
\sqrt{\frac{\omega}{2}}(\hat a - \hat a^{+}),  \nonumber \\
\hat n = a^+ a; \quad \bigl[a, a^+\bigr] = 1; \quad \hat n \left| n \right\rangle = n \left| n \right\rangle.
\end{eqnarray}

Some fairly simple algebraic operations with expressions (\ref{64}) - (\ref{69}) lead to the following results \cite{1}

\begin{eqnarray}
\label{72a}
&&E_{n}^{(0)}=\frac{1}{4\omega _{n}}(\omega _{n}^{2}+1+2\mu )(2n+1)+\frac{%
3\lambda }{4\omega _{n}^{2}}(1+2n+2n^{2});  \nonumber \\
&&\omega _{n}^{3}-\omega _{n}(1+2\mu )-\frac{6\lambda (2n^{2}+2n+1)}{2n+1}=0;
\end{eqnarray}

\begin{eqnarray}
\label{72b} &&\Delta E_{n}^{(2)}=-\frac{(n+1)(n+2)[\omega _{n}(1+2\mu -\omega _{n}^{2})+2\lambda
(2n+3)]^{2}}{16\omega _{n}^{2}[\omega _{n}(\omega _{n}^{2}+1+2\mu )+3\lambda
(3+2n)]}-\nonumber \\
&&\quad-  \frac{\lambda ^{2}(n+1)(n+2)(n+3)(n+4)}{32\omega _{n}^{2}[\omega
_{n}(\omega _{n}^{2}+1+2\mu )+3\lambda (5+2n)]}, \nonumber \\
&&\Delta E_{n}^{(3)} = -\frac{1}{256 \omega_n^2} \left\{ \frac{4 \left[(n+1)(n+2)\right]^{3/2}\left[ 2(2n+3)\lambda + \omega_n (1+2\mu)- \omega_n^3 \right]^3}{\left[ 3(2n+3) \lambda + \omega_n(1+2\mu) + \omega_n^3  \right]^2}\right.+ \nonumber \\
&&\quad+ \frac{2 \lambda \left[(n+1)(n+2)\right]^{3/2}\sqrt{(n+3)(n+4)}\left[ 2(2n+3)\lambda + \omega_n(1+2\mu) - \omega_n^3 \right]^2}{\left[ 3(2n+3) \lambda + \omega_n(1+2\mu) +  \omega_n^3  \right]\left[ 3(2n+5) \lambda + \omega_n(1+2\mu) +  \omega_n^3  \right]} + \nonumber \\
&&\quad+ \frac{2 \lambda^2 (n+3)(n+4) \left[(n+1)(n+2)\right]^{3/2}\left[ 2(2n+3)\lambda + \omega_n(1+2\mu) - \omega_n^3 \right]}{\left[ 3(2n+3) \lambda + \omega_n(1+2\mu) +  \omega_n^3  \right]\left[ 3(2n+5) \lambda + \omega_n(1+2\mu) +  \omega_n^3  \right]} + \nonumber \\
&&\quad+ \left.\frac{\lambda^3\left[ (n+1)(n+2)(n+3)(n+4) \right]^{3/2}}{\left[ 3(2n+5) \lambda + \omega_n(1+2\mu) +  \omega_n^3  \right]^2}\right\}
\end{eqnarray}

It can be shown by strict comparison with numerical results \cite{Hioe} that simple analytical expressions
(\ref{72a}) satisfy both definitions of the USE with $\xi ^{(0)}\simeq 0.03$. Fig.1 and Fig.2 illustrate this
statement. They show the results of calculation of the QAO energy levels in dependence on the parameter
$\lambda $ and quantum number $n$, obtained by means of various methods: numerical calculations (scattered)
\cite{Hioe}; the CPT; expansion in the strong coupling limit; the algebraic formulae (\ref{72a}) and (\ref{72b}) in the OM
zeroth, second and third orders. The absolute values and relative errors of various calculation methods in dependence
on the parameter $\lambda $ are compared in Fig.1, and Fig.2 shows the relative error in dependence on the
quantum number $n$. One can see that only approximation for the functions $E_{n}(\lambda )$ obtained in terms
of the OM does satisfy the definition (\ref{1}) for the USE approximated energy levels of the QAO.

\begin{figure}[p]\label{figure1}
\includegraphics[width=15cm,height=10.5cm]{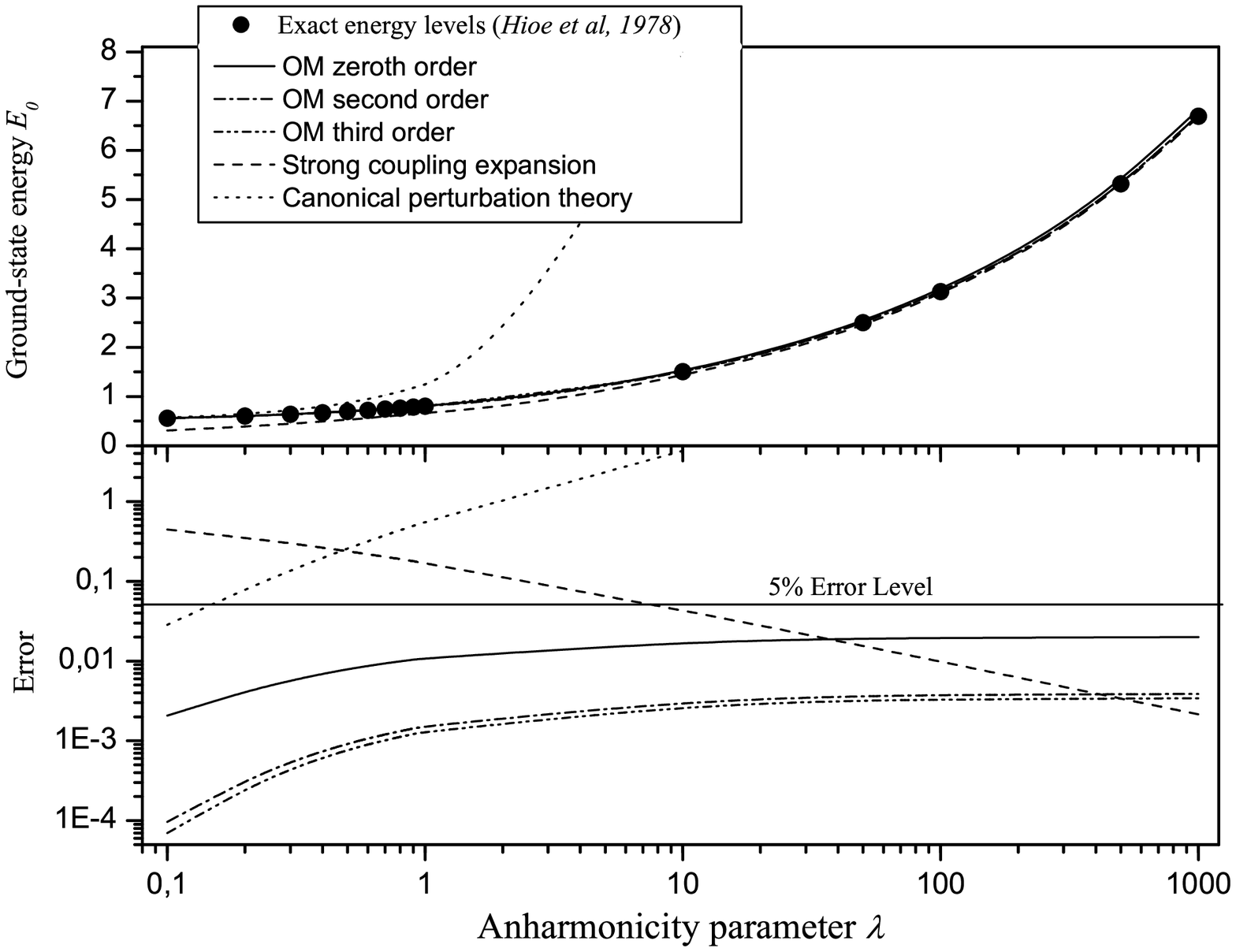}
\caption{Various estimations for the ground-state energy of the quantum anharmonic oscillator in dependence on
the anharmonicity parameter $\lambda $.}
\includegraphics[width=15cm,height=10.5cm]{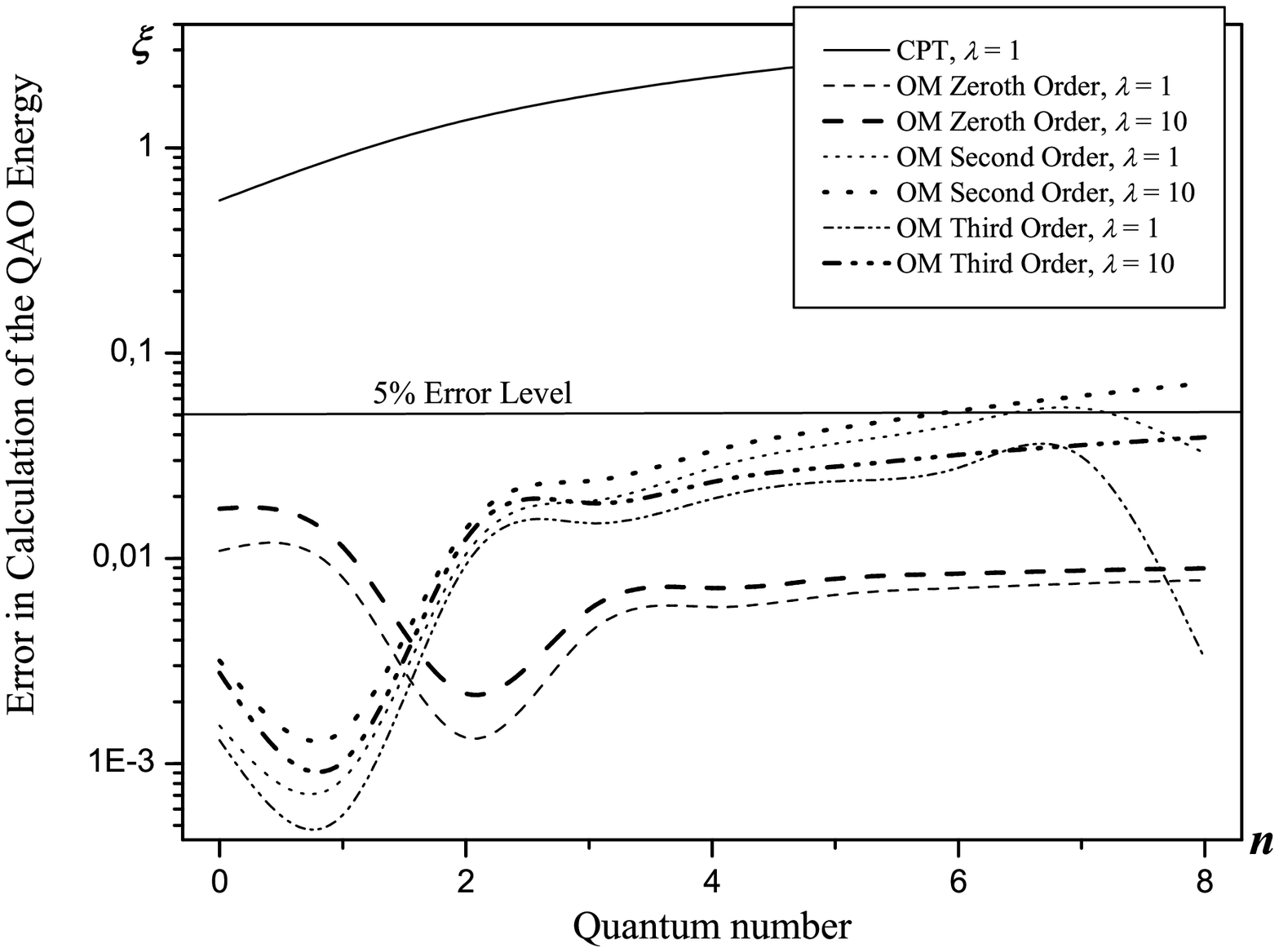}
\caption{Errors in calculation of the energy levels of the QAO in terms of various approximations considered as
the functions of quantum numbers.}

\end{figure}

\newpage

\section{Operator representation for the partition function and the cumulant
expansion}

Let us consider the partition function of some quantum system

\begin{equation}  \label{8a}
Z(\beta) = \sum_{n=0}^{\infty} g_n \exp[-\beta E_n],
\end{equation}

\noindent where $E_n$ is the energy level corresponding to a quantum number $%
n$ and degeneracy $g_n$.

To obtain the USE for this function one has to solve two problems: 1) to
construct the USE for the energy spectrum of a considered system that is valid for any quantum numbers, and 2) to obtain an
approximate estimation for summation on quantum numbers suitable in the
whole range of temperature. It should be pointed out, that for
single-dimensional systems the latter problem is less important. It is
sufficient to use expression (\ref{67}) for the energy levels, obtained in
terms of the OM, to derive the USE for thermodynamic values and to carry out
direct numerical summation (see below) in (\ref{8a}). However,
for systems with many degrees of freedom such summation is rather
complicated numerical problem, therefore we shall discuss the possibility of
constructing the USE for summation on quantum states.

First, let us show that the partition function can be identically
represented in the operator form as an average over some state vector. For
that purpose one can use the basic set of the state vectors considered
formally as the eigenfunctions of some excitation number operator

\begin{equation}  \label{8}
\hat n \left|n\right\rangle = n \left| n\right\rangle.
\end{equation}

Then one can rewrite equation (\ref{8a}) as

\begin{equation}  \label{9}
Z(\beta) = \left\langle \beta^* \right| \exp[-\beta E(\hat n) + 2 \beta^*
\hat n - \ln{N(\beta^*)}] \left| \beta^* \right\rangle.
\end{equation}

Here $\left\vert \beta ^{\ast }\right\rangle $ is the normalized
\textquotedblleft trial\textquotedblright\ state vector that depends on an
arbitrary parameter $\beta ^{\ast }$ having the physical meaning of an
effective inverse temperature for the equilibrium system of excitations
considered in (\ref{8}). The actual value of $\beta ^{\ast }$ will be
defined later from the condition of the best approximation for the partition
function of a real system. By definition, vector $\left\vert \beta ^{\ast
}\right\rangle $ is of  the following form

\begin{eqnarray}  \label{10}
\left| \beta^* \right\rangle = \sqrt{N} \sum_{n=0}^{\infty} \sqrt{g_n} \exp
[- \beta^* n]\left| n \right\rangle;  \nonumber \\
N = \left[\sum_{n=0}^{\infty} g_n \exp (- 2\beta^* n)\right]^{-1}.
\end{eqnarray}

It is important that the state vector $\left| \beta^* \right\rangle$ should not be considered as some
expression for the mixed state corresponding to Gibbs ensemble. Actually, equations (\ref{8})~-- (\ref{10}) give the identical representation for the numerical (non-operator) value (\ref{8a}) in the form
of quantum mechanical average using the auxiliary operator $\hat n$. It is also necessary to stress
that the operational representation for the partition function, analogous to (\ref{9}), can  be obtained by
choosing a more complicated trial vector instead of (\ref{10}). Such a vector should take into account
specific features of the problem under consideration. In every particular case, such a choice may lead to a
more accurate zeroth order approximation, but a simple vector (\ref{10}) used in this paper allows us to construct
the universal scheme for calculating the USE that is valid for an arbitrary quantum system.

Besides, this representation is formulated without usage of some specific form for the degeneracy factors $g_n$.
In general case the exact calculation of this factor in advance could prove to be  a rather complicated
problem. But when the OM is used for the approximate calculation of the energy levels $E_n$ the factor $g_n$
can be calculated with the same accuracy by means of approximate solution of equation (\ref{67a})
which defines the degeneracy of the state with the quantum number $n$ \cite{1}.

Now, we can use the cumulant expansion (CE) for calculating the average
value (\ref{9}). Recall that the CE is valid for an arbitrary exponential
operator when averaging over the normalized state vector \cite{2}

\begin{equation}
\left\langle \exp {\hat{A}}\right\rangle =\exp \left[ \sum_{n=1}^{\infty }%
\frac{K_{n}}{n!}\right] ,  \label{11}
\end{equation}

\noindent where the cumulants $K_{n}$ are expressed in terms of the moments
of the operator $\hat{A}$. This expansion as a whole is an accurate one but
at the same time every cumulant corresponds to the partial summation of a
usual power series. As follows from \cite{2}, the first few terms in
equation (\ref{11}) are given by

\begin{eqnarray}  \label{12}
K_1 = \left\langle \hat A\right\rangle; \quad K_2 = \left\langle \hat A^2
\right\rangle - \left\langle \hat A\right\rangle^2;  \nonumber \\
K_3 = \left\langle \hat A^3 \right\rangle - 3 \left\langle \hat A
\right\rangle \left\langle \hat A^2 \right\rangle + 2\left\langle \hat A
\right\rangle^3.
\end{eqnarray}

The general procedure of the CE can be applied to the partition function in form (\ref{9}). Then successive
approximations for the partition function will be calculated taking into account the corresponding cumulants.
In this paper, we restrict ourselves by the first and second cumulants which leads to the following
approximation

\begin{eqnarray}  \label{13}
Z \simeq Z(\beta, \beta^*) = Z_0(\beta, \beta^*)Z_1(\beta, \beta^*) =
\nonumber \\
\exp \left[ \left\langle A(\hat n)\right\rangle - \ln N + \frac{1}{2} \left(
\left\langle A^2(\hat n)\right\rangle - \left\langle A(\hat n)
\right\rangle^2 \right) \right];  \nonumber \\
A(\hat n) = -\beta E(\hat n) + 2\beta^* \hat n.
\end{eqnarray}

It is obvious that with any fixed number of cumulants in formula (\ref{13})
the partition function $Z(\beta ,\beta ^{\ast })$ depends on the artificial
value $\beta ^{\ast }$, which can be considered as a variational parameter,
leading to the best approximation of the CE in some order. Particularly, the
zeroth order approximation $Z_{0}(\beta ,\beta ^{\ast })$ and the first correction
$Z_{1}(\beta ,\beta ^{\ast })$ are defined by formulae

\begin{eqnarray}  \label{14}
Z_0 = \exp \left[ -\beta \bar E(\beta^*) + 2\beta^* \bar n - \ln N
(\beta^*)\right ];  \nonumber \\
Z_1 = \exp \{\frac{1}{2}\left[ \beta^2 (\overline{E^2} - \bar{E}^2) - 4\beta
\beta^* (\overline{E n} - \bar{E}\bar{n}) + 4\beta^{*2}(\overline{n^2} -
\bar n^2)\right]\}.
\end{eqnarray}

Here, all values are averaged over the \textquotedblleft
trial\textquotedblright\ distribution function corresponding to the ensemble
of the excitations (\ref{8}) with the degeneracy of states $g_{n}$ and the
effective inverse temperature $\beta ^{\ast }$, as for example

\begin{eqnarray}  \label{15}
\bar E (\beta^*) = N \sum_{n=0}^{\infty} g_n E_n \exp [- 2\beta^* n];
\nonumber \\
\bar n(\beta^*) = N \sum_{n=0}^{\infty} g_n n \exp [- 2\beta^* n] = \frac{N}{%
2}\frac{\partial N}{\partial \beta^*}.
\end{eqnarray}

Then the free energy calculated with the considered accuracy takes the
following form

\begin{equation}  \label{16}
F(\beta)= -\frac{1}{\beta} \ln Z(\beta) \simeq F^{(0)} + F^{(1)} + \cdots = -%
\frac{1}{\beta} (\ln Z_0[\beta^*,\beta] + \ln Z_1[\beta^*,\beta] + \cdots).
\end{equation}

To obtain the estimation for the partition function of the given energy
spectrum, the variational parameter $\beta ^{\ast }$ considered as the
function of the real temperature $\beta $ can be defined from the condition
of the best approximation in the zeroth order of the CE. As follows from (%
\ref{69}), for this purpose one has to solve the equation

\begin{equation}  \label{161}
\frac{\partial Z_0}{\partial \beta^*}=0; \; \Rightarrow \; \beta \frac{%
\partial \bar E}{\partial \beta^*} - 2\bar n - 2\beta^*\frac{\partial \bar n%
}{\partial \beta^*} + \frac{\partial N}{N\partial \beta^*}=0.
\end{equation}

One can use the energy spectrum, obtained in terms of the OM to calculate
the moments $\bar{E},\overline{E^{2}}$ with the distribution function (\ref%
{15}). Then the totality of expressions (\ref{67}), (\ref{69}), (\ref{14}%
) and (\ref{161}) lead to the USE for the thermodynamic characteristics of a
system with the accuracy to the terms of second order in approximation of
the OM and the CE. In some cases, the additional approximation in these
equations can also be applied. This approximation leads to the worst
accuracy, but permits one to simplify significantly the evaluations. It is
based on the following.

Recall that condition (\ref{69}) corresponds to the optimal choice of the parameter $\omega _{n}$ for the best
approximation for the energy level with quantum number $n$ in the zeroth order of the OM. But convergence
of the successive approximations of the OM exists for all values of $\omega $ \cite{1}. At the same time, the
partition function and the free energy are integral characteristics in the space of quantum numbers of a
system. Thus, it is possible to expect that in order to construct the USE for these values it is sufficient to
choose the parameter $\omega$ as the same value for all quantum numbers, and to use it for defining the best
approximation condition for the free energy in the zeroth order of the OM and the CE. This means that for the
calculation of the moments of the energy spectrum one has to take into account only the explicit dependence of
the approximate energy spectrum on the quantum number $n$ and disregard a more complicated dependence related
to the choice of $\omega _{n}$ from condition (\ref{69}). Hence, the values

\begin{eqnarray}  \label{162}
\bar E (\beta^*,\omega) \simeq N \sum_{n=0}^{\infty} g_n [E^{(0)}_n (\omega)
+ \Delta E_n (\omega)] \exp [- 2\beta^* n];  \nonumber \\
\overline{E^2}(\beta^*,\omega) \simeq N \sum_{n=0}^{\infty} g_n [E^{(0)}_n
(\omega)]^2 \exp [- 2\beta^* n]
\end{eqnarray}

\noindent depend on both the parameter $\beta ^{\ast }$ and the variational parameter $\omega $, which should be
taken from the minimum condition for the free energy. As a result, instead of a system of algebraic equations
(\ref{14}) for $\omega _{n}$ it is necessary to solve only one equation for the parameter $\omega $

\begin{eqnarray}  \label{163}
\frac{\partial}{\partial\omega} \sum_{n=0}^{\infty} g_n [E^{(0)}_n
(\omega)]\exp [- 2\beta^* n] = 0.
\end{eqnarray}

The physical meaning of this equation is that the variational parameter $%
\omega $ should be chosen from the condition of the best approximation of
the energy level with maximum occupancy corresponding to a given temperature.

In this paper, we illustrate the efficiency of our method in obtaining the USE for some specific systems. For
these systems, all the calculations can be performed analytically. In spite of the model pattern of the
problems under consideration, they are rather widely used for investigations of thermodynamic properties of
real molecular gases \cite{Gribov}, therefore, the set of analytical formulae considered below may be useful
for nonperturbative analysis of anharmonic effects in such systems.

It should be pointed out that in the case of harmonic oscillator ($\lambda =0
$) in (\ref{71}), rather simple calculations show that, unlike the  CPT (\ref%
{6}), formulae (\ref{162}),(\ref{163}) lead to the exact expression for the
partition function with any value of parameter $\mu $.

As mentioned above, the proposed method of calculating the USE for the partition function includes actually two
components: the approximate summation over the quantum states and the approximate calculation of the energy
spectrum. In order to illustrate how the first one works, let us obtain the USE for the partition function and
the free energy for the system of quantum rotators. Such a system is used for simulation of the contribution of
rotational degrees of freedom to the thermodynamic characteristics of diatomic molecular gases \cite{5}. We can
consider this system as a model with known energy spectrum and degeneracy, therefore it can be considered as
a test for approbation of the cumulant expansion for approximate summation over the states with known
spectrum. Hence, we should calculate the following partition function \cite{5}

\begin{equation}  \label{18}
Z_r(\beta, \theta_r) \equiv Z_r(x) = \sum_{n=0}^{\infty}(2n + 1)\exp [-\beta
\theta_r(n^2 + n)].
\end{equation}

Here $\theta _{r}=\hbar ^{2}/2I$ is the rotational temperature of the system
measured in the energy units; $I$ is the moment of inertia of the considered
molecules. Let us introduce $x$ as the dimensionless
parameter, $x=\beta \theta _{r}$.

One can find analytically the asymptotic formulae for small and large
temperatures which are connected with the limit cases $x \gg 1$ and $x \ll 1$
correspondingly

\begin{eqnarray}  \label{19}
Z_r(x) \simeq 1 + 3 \exp (- 2 x) + \ldots; \qquad x \gg 1.
\end{eqnarray}

In the opposite limit case, we use two terms of the Euler formula of
summation
\[
\sum_{n=1}^{\infty }f(n+a)\simeq \int_{a}^{\infty }f(y)dy-\frac{1}{2}f(a)-%
\frac{1}{12}f^{\prime }(a),
\]%
which leads to

\begin{eqnarray}  \label{20}
Z_r(x) \simeq 1 + \int_0^{\infty} (2y+1)\exp [- x(y^2 + y)] dy - \frac{1}{2}
- \frac{1}{12}(2 - x) + \ldots =  \nonumber \\
= \frac{1}{x} + \frac{1}{3} + \frac{x}{12}; \qquad x \ll 1.
\end{eqnarray}

The presented expressions do not describe adequately the function $Z_r(x)$ in the range of intermediate
values of the parameter $x$. Actually $Z_r(x)$ can be expressed through the Weierstrass function \cite{Stigan},
but it is rather complicated for analytical investigations. Let us show that the usage of the proposed method
allows one to obtain the USE for this function in the entire range of $x$.

In such a case all average values in formula (\ref{15}) can be directly
expressed through moments of the ``trial'' distribution function

\begin{eqnarray}  \label{21}
N = \left[\sum_{n=0}^{\infty}(2n + 1) e^{-2\beta^*n} \right]^{-1} = \frac{(1
- q)^2}{1 + q}; \quad q = e^{-2\beta^*};  \nonumber \\
\bar n = N \sum_{n=0}^{\infty}n(2n + 1) e^{-2\beta^*n} = \frac{q (3 + q)}{1
- q^2};  \nonumber \\
\overline{n^2}= \frac{q(3 + 8 q + q^2)}{(1+q)(1-q)^2}.
\end{eqnarray}

Using the presented formulae in (\ref{14}), we can obtain the following
expression for the partition function in the zeroth order of the CE (the
value $0 < q < 1$ can be considered as a variational parameter instead of $%
\beta^*$)

\begin{eqnarray}  \label{22}
Z^{(0)}_r(x,q) = \exp [\varphi (x, q )];  \nonumber \\
\varphi (x, q ) = - x \frac{6q}{(1-q)^2} - \frac{3q + q^2}{1 - q^2}\ln q -
\ln \left[\frac{(1 - q)^2}{1 + q} \right].
\end{eqnarray}

This formula should be supplemented with an equation for calculating the
function $q = q(x)$, following from condition (\ref{161}) for the best
choice of the zeroth order approximation

\begin{eqnarray}  \label{23}
\frac{\partial \varphi}{\partial q} = 0; \quad x = -\frac{(3 + 2q +
3q^2)(1-q)}{6(1 + q)^3}\ln q = 0.
\end{eqnarray}

\begin{eqnarray}  \label{231}
Z^{(0)}_r(q) = \exp [\varphi (q )];  \nonumber \\
\varphi (q ) = q \left[ \frac{3 + 2 q + 3 q^2}{(1 - q)(1 + q)^2} - ( 3 +
q)(1 - q^2) \right]\ln q - \ln \left[ \frac{(1 - q)^2}{1 + q} \right].
\end{eqnarray}

System of equations (\ref{23}), (\ref{231}) should be considered as a
parametric assignment of the function $Z^{(0)}_r(x)$, defining the zeroth order
approximation for the partition function.

The first order correction to the CE for the considered model spectrum is
defined by the expression

\begin{eqnarray}
Z_{r}^{(1)}(x,q) &=&\exp [\varphi _{1}(x,q)];  \nonumber  \label{24} \\
\varphi _{1}(x,q) &=&\frac{1}{2}\ x^{2}\left[ \overline{(n^{2}+n)^{2}}-%
\overline{(n^{2}+n)}^{2}\right] +2x\left[ \overline{n(n^{2}+n)}-\overline{%
(n^{2}+n)}\bar{n}\right] \ln q+\left[ \overline{n^{2}}-\bar{n}^{2}\right]
\ln ^{2}q.
\end{eqnarray}

Using equation (\ref{23}) in the previous expression, one can find the
following

\begin{eqnarray}  \label{25}
Z^{(1)}_r(q) = \exp [\varphi_1 (q )];  \nonumber \\
\varphi_1 ( q ) = \frac{q^2}{6}\frac{15 + 4 q + 26 q^2 + 4 q^3 + 15 q^4}{(1
- q)^2 (1 + q)^6} \ln^2 q.
\end{eqnarray}

The presented formulae define the USE for the partition function of the
quantum rotator. In particular, they lead to the following results in the
corresponding limit cases

\begin{eqnarray}  \label{26}
x \simeq - \frac{\ln q}{2}, \quad \varphi (q) \simeq 3 q, \quad Z^{(0)}_r(x)
\simeq 1 + 3 \exp (- 2 x) + \ldots, \quad x \gg 1;  \nonumber \\
x \simeq \frac{(1-q)^2}{6}, \quad \varphi (q) \simeq 1 + \ln 2 - 2 \ln (1 -
q),\quad Z^{(0)}_r(x) \simeq \frac{0.906}{x} + 0.303 + \ldots, \quad x \ll 1.
\end{eqnarray}

Comparing these results with corresponding asymptotic formulae (\ref{19}), (%
\ref{20}), one can see, that the constructed approximation describes
correctly the functional dependence of the exact partition function in both
limit cases. At the same time in compliance with the common definition (\ref%
{1}) for the USE the accuracy of the estimation is defined by parameter $%
\xi^{(0)} \simeq 0.1$. The first-order correction (\ref{25}) of the CE
increases the accuracy of the estimation, in particular, when the value of $%
x $ is small

\begin{eqnarray}  \label{27}
Z_r(x) \simeq Z^{(0)}_r(x) Z^{(1)}_r(x)\simeq \frac{1.063}{x} + 0.348 +
\ldots, \quad x \ll 1.
\end{eqnarray}

Fig.3 represents the comparison of the results of evaluations in terms of
exact and approximate analytical formulae in the zeroth and second orders of
the CE. It demonstrates rather good accuracy of the zeroth order approximation of
the CE and its uniform validity for all values of the dimensionless
parameter, defining the properties of the discussed system.

\begin{figure}[ht]\label{figure2}
\includegraphics[width=15cm,height=11cm]{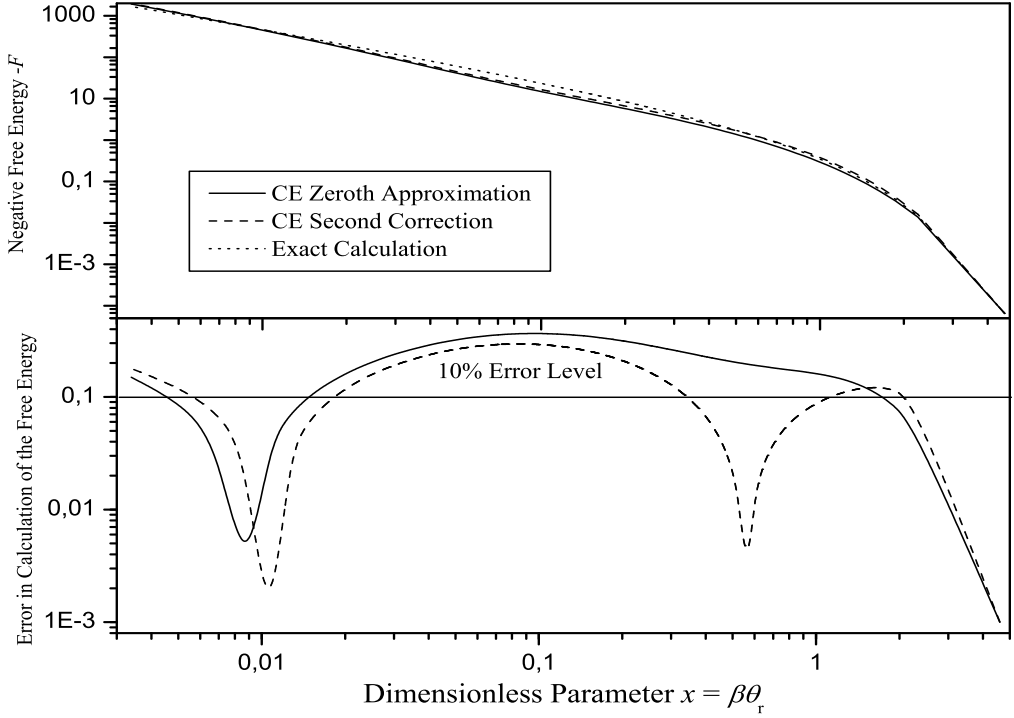}
\caption{Free energy of quantum rotator and error in its calculation in terms of the CE in zeroth and second order in dependence on the dimensionless parameter $x$.}
\end{figure}

\newpage

\section{Thermodynamics of the Quantum Anharmonic Oscillator}

The problem considered in this section illustrates the possibilities of the
discussed method in the case when the approximate
calculations are used for obtaining the USE for both the eigenvalues and
summation over the states. For that purpose, we use the equilibrium
statistical system of quantum anharmonic oscillators (QAO). This problem
has been discussed in numerous papers concerning nonperturbative approximate
methods (e.g. \cite{7} - \cite{16}), but the approximation satisfying
definitions (\ref{1}) and (\ref{2}) has not been obtained yet. As
mentioned above, the solution of this problem can be useful for
applications, in which one should take into account the anharmonicity of the
molecular oscillations (for example, when considering either the
thermodynamics of gases \cite{Gribov} or the anharmonic effects in
calculations of phonon spectra in crystals \cite{2d}).

Let us consider the QAO with parameter $\mu =0$. This can be done without
loss of generality, since the parameter $\mu $ can always be excluded
from the Hamiltonian by the scaling transformation of the coordinate. We thus consider

\begin{equation}  \label{28}
\hat H = \frac{1}{2} (\hat p^2 + \hat x^2) + \lambda \hat x^4.
\end{equation}

The USE for this system is defined by formulae (\ref{72a}) and (\ref{72b}). When obtaining the
USE for the partition function and the free energy it is interesting to
compare the results, which can be obtained by means of approximate
calculations in two different ways: 1) direct numerical summation over the
QAO states taking into account the OM energy levels and 2) evaluation of the
USE in analytical form by means of the approximate summation over the states
based on formulae (\ref{162}), (\ref{163}). To estimate the accuracy of our
results we will use direct numerical calculation of the partition function (%
\ref{8a}) ($Z_{A}^{num}$) for the QAO ($g_{n}=1$) taking into account 8 energy levels of the QAO, obtained
numerically in paper \cite{Hioe}. It is essential to stress, that in real calculations with any fixed number of
the QAO levels in the summation over the states, the numerical calculations lose their accuracy in the limit
case of the high temperature ($\beta \rightarrow 0$), when the whole energy spectrum becomes essential (see
left panel of Fig.4). In these cases one can use for comparison the asymptotic expressions for the partition function which
can be obtained analytically:

\begin{eqnarray}  \label{29}
Z_A(\beta, \lambda) = \sum_{n=0}^{\infty} \exp[-\beta E_n(\lambda)] \simeq 1
+ e^{-\beta E_0(\lambda)}, \quad \beta \gg 1;  \nonumber \\
Z_A(\beta, \lambda) \simeq \frac{e^{-\beta/2}}{1 - e^{-\beta}} \left[ 1 +
\frac{3\lambda e^{-\beta}}{(1 - e^{-\beta})^2}\right], \quad \frac{\lambda }{%
(1 - e^{-\beta})^2} \ll 1;  \nonumber \\
Z_A(\beta, \lambda) \simeq \sum_{n=0}^{\infty} \exp[-\beta \lambda^{1/3} b_n]%
, \quad \lambda \gg 1.
\end{eqnarray}

Here $b_{n}$ are known numerical coefficients of the asymptotic expansion
for the QAO energy levels in the strong coupling limit \cite{Hioe}. The
analytical approximate expression for them in the OM zeroth order was
obtained in paper \cite{1}

\begin{equation}  \label{30}
b_n \simeq (\frac{3}{4})^{4/3}[\frac{1 + 2n + 2n^2}{(1 + 2n )^2}]^{1/3}.
\end{equation}

Fig.4 represents the results of comparison of the free energy $%
F_{A}^{num}=-1/\beta \ln (Z_{A}^{num})$ with various approximate expressions for it, which are defined as the
following

\begin{eqnarray}  \label{31}
F^{(0)}_A(\beta, \lambda) = -\frac{1}{\beta} \ln \left[ \sum_{n=0}^{\infty}
\exp[-\beta E^{(0)}_n(\lambda)] \right];  \nonumber \\
F^{(1)}_A(\beta, \lambda) = -\frac{1}{\beta} \ln \left[ \sum_{n=0}^{\infty}
\exp[-\beta (E^{(0)}_n + \Delta E_n)] \right];  \nonumber \\
F^{(01)}_A(\beta, \lambda) = -\frac{1}{\beta} \ln \left[ \sum_{n=0}^{\infty}
\exp[-\beta E^{(0)}_n](1 - \beta \Delta E_n)\right],
\end{eqnarray}

\noindent where the energy is calculated in different orders of the OM by formulae (\ref{72a}), (\ref{72b}). We
also compare our approximations with the results of the canonical thermodynamic perturbation theory
\cite{5} and the direct summation over the exact eigenvalues for the first 8 levels of QAO calculated
numerically in \cite{Hioe}.

\begin{figure}[ht]\label{figure3}
\includegraphics[width=18.5cm,height=13cm]{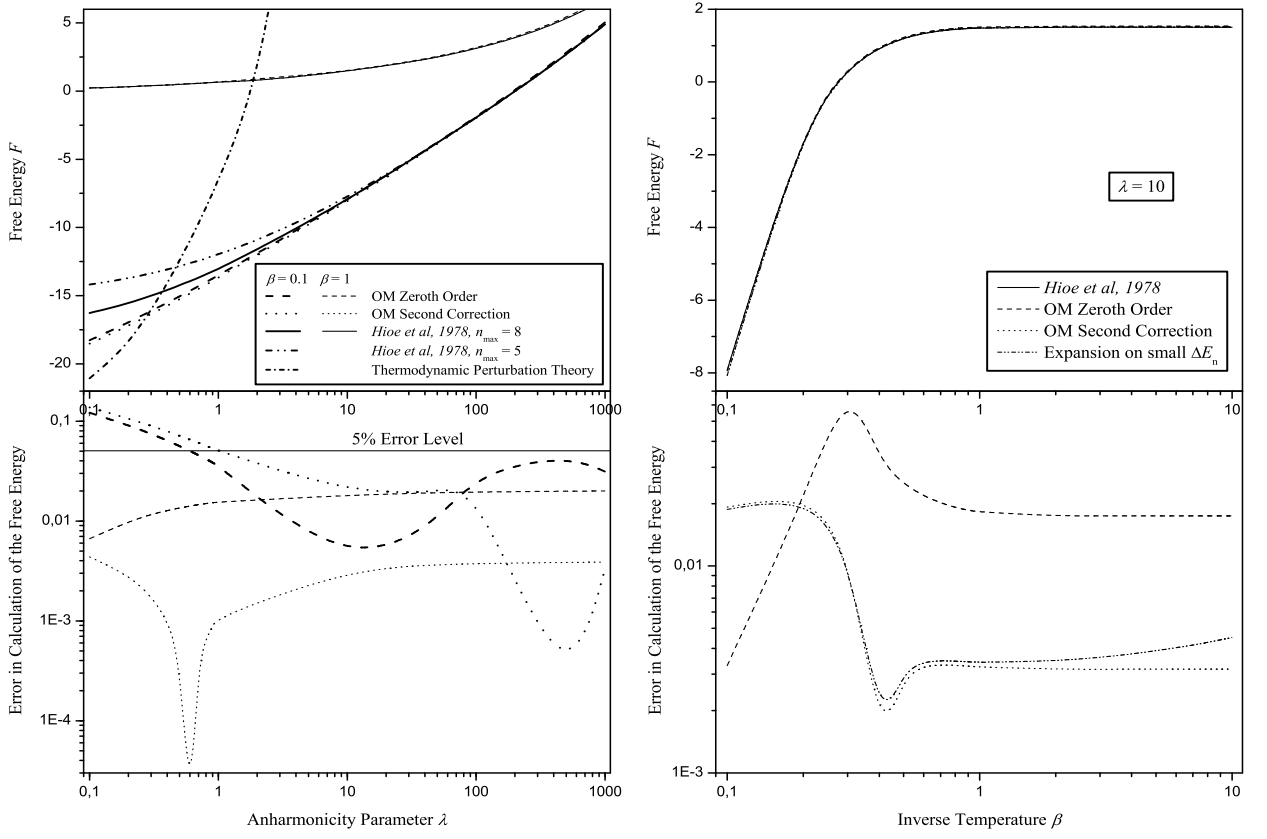}
\caption{The free energy of the QAO using the OM zeroth and second orders and errors of these approximations in
dependence on the anharmonicity parameter $\lambda$ (left panel) and on the inverse temperature $\beta$ (right
panel) in comparison with the results of the numerical calculations. Lines, corresponding to $\beta = 1$ on the
upper part of the left panel, and all lines on the upper part of the right panel almost coincide. Results
obtained in terms of the thermodynamic perturbation theory  are also represented on the upper part of the left
panel (they are not uniformly suitable and can not be shown on the right panel).}
\end{figure}

The presented results show that the zeroth order approximation $F_{A}^{(0)}$ satisfies condition (\ref{1}) and
defines the USE for the pointed values with a relative error $\xi ^{(0)}\simeq 0.1$. If one takes into account the
subsequent iteration of the OM, the accuracy of the estimation improves. At the same time, it is necessary to
stress that the expansion of the exponent in a rather small value $\Delta E_{n}$ used in expression
$Z_{A}^{(01)}(\beta ,\lambda )$ breaks the USE conditions in the low temperature limit ($\beta \rightarrow
\infty $). This means that the improvement of the accuracy of the zeroth order approximation for the free energy in
the subsequent iterations of the considered method has in general nonadditive character, unlike the
thermodynamic CPT \cite{5}.

Let us consider now the approximate expression for the partition function
obtained in terms of the zeroth order approximation of both the OM and the CE due to
formulae (\ref{161}) - (\ref{163})

\begin{eqnarray}  \label{32}
Z^{(0C)}_A(\beta, \lambda) = \exp[\varphi(\beta, \beta^*,\lambda, \omega)];
\nonumber \\
\varphi(\beta, \beta^*,\lambda, \omega) = - \beta \overline{%
E_n^{(0)}(\beta^*,\omega,\lambda)} + 2\beta^*\bar n (\beta^*) - \ln N;
\nonumber \\
N = (1 - q); \bar n = \frac{q}{(1- q)}; \overline{n^2}= \frac{q + q^2}{(1 -
q)^2}; q = e^{- 2 \beta^*};  \nonumber \\
\overline{E_n^{(0)}} = \frac{1}{4 \omega} (\omega^2 +1) (2 \bar n + 1) +
\frac{3\lambda}{4 \omega^2} (1 + 2 \bar n + 2 \overline{n^2}).
\end{eqnarray}

Variational parameters $q$ and $\omega $ are defined from the minimum conditions for function $\varphi $, which lead to the following equations

\begin{eqnarray}  \label{33}
\frac{\partial \varphi}{\partial q}=0, \; \Rightarrow \; -\beta \frac{%
\omega^2 + 1}{2 \omega (1 - q)} + \frac{3 \lambda (1 + q)}{\omega^2 (1 - q)^2%
} + \frac{2 \ln q}{1 - q} + 1 = 0;  \nonumber \\
\frac{\partial \varphi}{\partial \omega}=0, \; \Rightarrow \; \omega^3 -
\omega - 6 \lambda \frac{1 + q}{1 - q} = 0.
\end{eqnarray}

It is convenient not to solve numerically equations (\ref{33}), but to consider $\lambda (q,\omega )$ and $\beta
(q,\omega )$ as functions of variables $q$ and $\omega $. Together with definition (\ref{32}) these functions
assign in parametric form the functions $Z_{A}^{(0C)}(\beta ,\lambda )$ and $F_{A}^{(0C)}(\beta ,\lambda
)=-1/\beta \ln Z_{A}^{(0C)}(\beta ,\lambda )$. The results of such a calculation and its comparison with the
numerically calculated free energy are shown in Fig.5. As one can see, the application of the CE to summation
over the states keeps the zeroth order approximation of the method under consideration satisfying the criteria
of the USE with a relative error $\xi ^{(0)}\simeq 0.1$. Fig.5 shows also that the values $Z_{A}^{(0C)}(\beta
,\lambda )$ and $F_{A}^{(0C)}(\beta ,\lambda )$ are rather close to the corresponding values $Z^{(0)}_A(\beta,
\lambda)$ and $F^{(0)}_A(\beta, \lambda)$ calculated by means of direct summation over the OM eigenvalues
$E^{(0)}_n (\omega_n)$. In the latter case the parameter $\omega_n$ should be defined for each level separately.
The former approach based on equations (33), (34) is much easier because the parameter $\omega$ is
calculated only once for each temperature. It is especially important for systems with many degrees of
freedom.

To obtain the next approximation for the partition function $%
Z_{A}^{(1C)}(\beta ,\lambda )$ and the free energy $F_{A}^{(1C)}(\beta
,\lambda )=-1/\beta \ln Z_{A}^{(1C)}(\beta ,\lambda )$ it is necessary to
take into account both the correction $\Delta E_{n}$ for the energy levels
obtained in terms of the OM, and the second cumulant in formula (\ref{14})
for the approximate summation over the states. It is essential that in
calculation of these corrections the same variational parameters $q$ and $%
\omega $ are used, which were found in the zeroth order. Fig.6 demonstrates the influence of these corrections on the accuracy of the
USE.

\newpage

\begin{figure}[ht]\label{figure4}
\includegraphics[width=18.5cm,height=13cm]{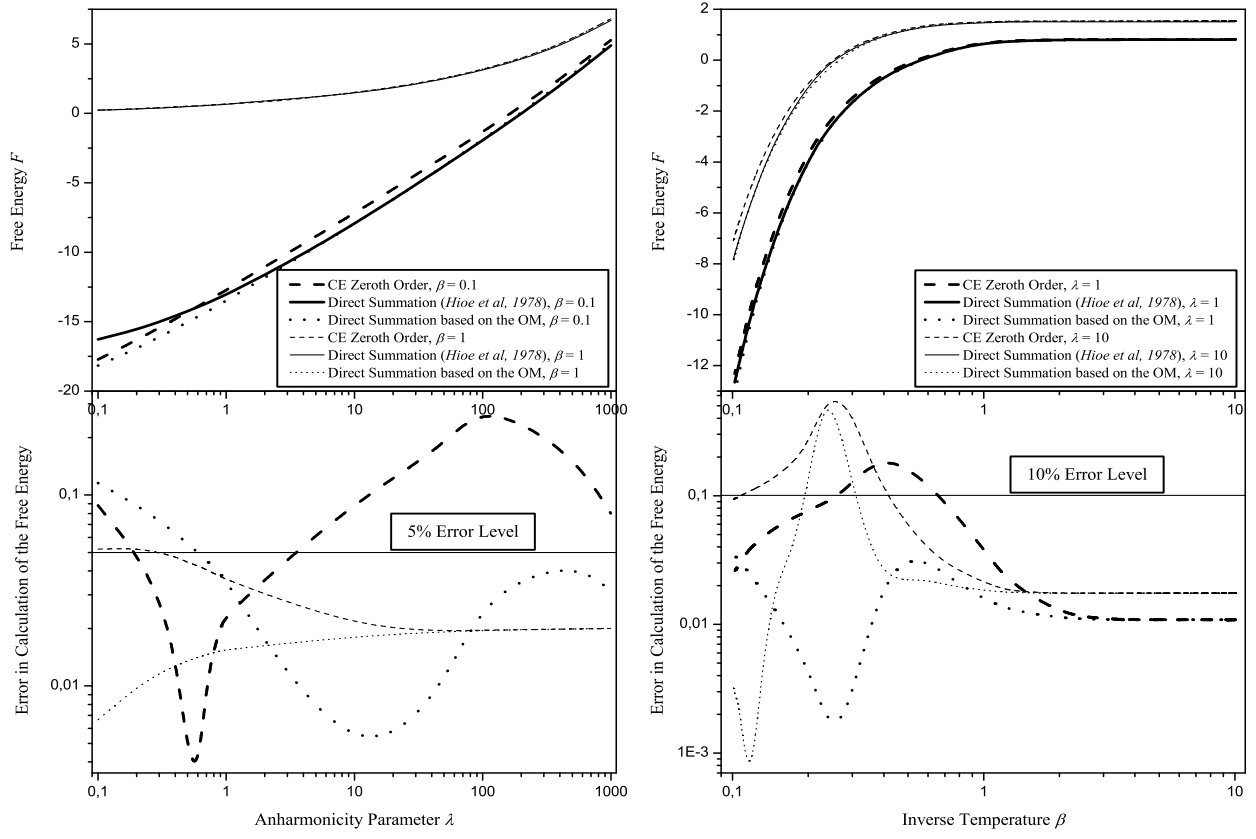}
\caption{Approximation for the free energy of the QAO using the CE zeroth order, direct numerical summation based on the OM zeroth order and exact results \cite{Hioe} together with the relative errors of such approximations in dependence on the anharmonicity parameter $\lambda$
for various values of inverse temperature $\beta$ (left panel) and on the inverse temperature $\beta$ for
various values of the anharmonicity parameter $\lambda$ (right panel). Lines 4, 5 and 6 on the upper part of the
left panel almost coincide.}
\end{figure}

\newpage

\begin{figure}[ht]\label{figure5}
\includegraphics[width=18.5cm,height=13cm]{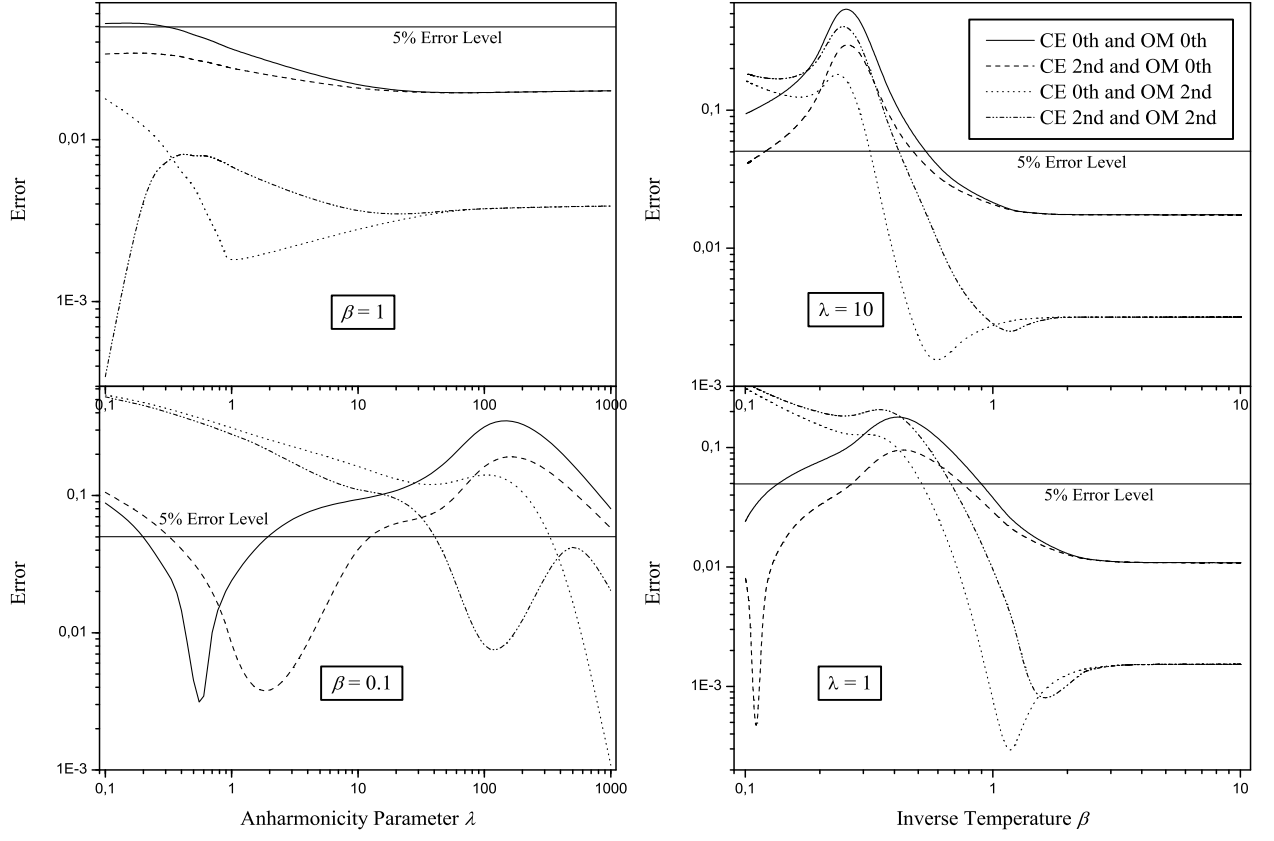}
\caption{Errors of various approximations for the free energy of QAO in dependence on the anharmonicity parameter $\lambda$ for various values of inverse temperature $\beta$ (left panel) and on the inverse temperature
$\beta$ for various values of the anharmonicity parameter $\lambda$ (right panel).}
\end{figure}

It is also of interest to consider the average values of observable physical characteristics of the system
(e.g., energy) in dependence on the Hamiltonian parameters and temperature. For this purpose we use the
numerical energy levels \cite{Hioe}, the energy levels obtained in terms of the CPT and the OM in zeroth order.
We consider also the expression for the average value of energy based on the cumulant expansion. Thus we obtain
the following expressions for the average energy

\begin{eqnarray}
\overline{E^{(num)}} &=& \frac{1}{Z_A^{(num)}}\sum_{n} E_n^{(num)}\exp{[-\beta E_n^{(num)}]}, \nonumber \\
\overline{E^{(0)}} &=& \frac{1}{Z_A^{(0)}}\sum_{n} E_n^{(0)}\exp{[-\beta E_n^{(0)}]}, \nonumber \\
\overline{E^{(CPT)}} &=& \frac{1}{Z_A^{(CPT)}}\sum_{n} E_n^{(CPT)}\exp{[-\beta E_n^{(CPT)}]}, \nonumber \\
\overline{E^{(CE)}} &=& -\frac{1}{Z^{(0C)}} \frac{\partial}{\partial \beta} Z^{(0C)}(\beta).
\end{eqnarray}

The results of these calculations are presented in Fig.7 which shows that our approach leads to the uniformly
suitable estimation also for the average physical values unlike the canonical perturbation theory. As it was
mentioned above, the noticeable deviation of the OM results from the numerical ones in the range of small
$\beta$ is explained by the fact that we could use only 8 eigenvalues found in \cite{Hioe} for calculation of
$\overline{E^{(num)}}$. It is also important to stress again that the value $\overline{E^{(0)}}$ calculated with
the OM eigenvalues depending on the set of the parameters $\omega_n$ is rather close to the value
$\overline{E^{(CE)}}$ depending on the only parameter $\omega$ from the equation (34).

\begin{figure}[ht]\label{figure7}
\includegraphics[width=18.5cm,height=13cm]{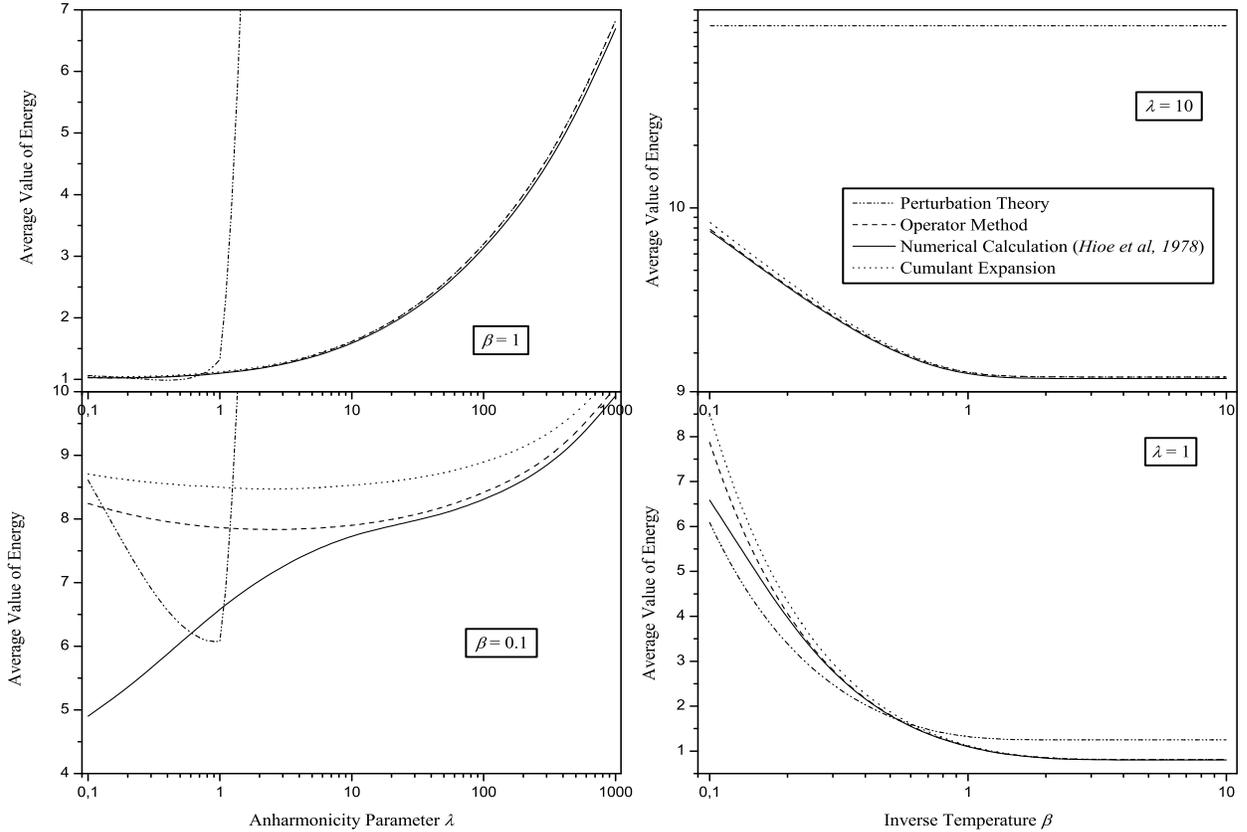}
\caption{Average values of the energy of the QAO in dependence on the anharmonicity parameter $\lambda$ for
various values of the inverse temperature $\beta$ (left panel) and on the inverse temperature $\beta$ for
various values of the anharmonicity parameter $\lambda$ (right panel).}
\end{figure}

\section{Conclusions}

In this paper, we have developed  a nonperturbative method for calculation of the thermodynamic values of a
quantum system. This has been achieved by combining the operator method of approximate solution of
Schr\"{o}dinger equation and the cumulant expansion for the summation over the quantum states. The method has
been approved for the Boltzmann diatomic molecular gas in order to calculate the partition function and the free
energy defined by the molecular internal degrees of freedom. We have used some realistic models for the
molecular movement (a quantum rotator and a quantum anharmonic oscillator) and found the uniformly suitable
estimation for the thermodynamic values. This estimation tends asymptotically to the exact expansions in limit
cases of both temperature and the Hamiltonian parameters. Besides, the zeroth order approximation of the
proposed method is in close agreement with the exact results (with the relative error no more than 0.1) in the whole
range of both temperature and the Hamiltonian parameters. A systematic procedure for calculation of the
subsequent corrections has been formulated and the second order corrections have been found to improve the
accuracy of the estimation. It has been also shown that application of the cumulant expansion for the summation
over the system states permits one to calculate directly the thermodynamic values without preliminary high
precision estimation of the whole set of the Hamiltonian eigenvalues. It is especially important for application
of the proposed algorithm for the systems with many degrees of freedom.

Certainly, the strict  proof of convergence of the formulated method is of special interest. Unfortunately it
seems to be a very complicated mathematical problem in general case. But approbation of the method for a series
of model systems in this paper can be considered as a qualitative argument for possibility of applying of this
approach to real physical problems which we suppose to analyse in the forthcoming papers.

\vspace{3mm} \hrule 
\newpage

\section{Captions for figures}

Figure 1. Various estimations for the ground-state energy of the quantum anharmonic oscillator in dependence on
the anharmonicity parameter $\lambda $.

Figure 2. Errors in calculation of the energy levels of the QAO in terms of various approximations considered
as the functions of quantum numbers.

Figure 3. Free energy of quantum rotator and error in its calculation in terms of the CE in zeroth and second
order in dependence on the dimensionless parameter $x$.

Figure 4. The free energy of the QAO using the OM zeroth and second orders and errors of these approximations in
dependence on the anharmonicity parameter $\lambda$ (left panel) and on the inverse temperature $\beta$ (right
panel) in comparison with the results of the numerical calculations. Lines, corresponding to $\beta = 1$ on the
upper part of the left panel, and all lines on the upper part of the right panel almost coincide. Results
obtained in terms of the thermodynamic perturbation theory  are also represented on the upper part of the left
panel (they are not uniformly suitable and can not be shown on the right panel).

Figure 5. Approximation for the free energy of the QAO using the CE zeroth order, direct numerical summation based on the OM zeroth order and exact results \cite{Hioe} together with the relative errors of such approximations in dependence on the anharmonicity parameter $\lambda$
for various values of inverse temperature $\beta$ (left panel) and on the inverse temperature $\beta$ for
various values of the anharmonicity parameter $\lambda$ (right panel). Lines 4, 5 and 6 on the upper part of the
left panel almost coincide.

Figure 6. Errors of various approximations for the free energy of QAO in dependence on the anharmonicity
parameter $\lambda$ for various values of inverse temperature $\beta$ (left panel) and on the inverse
temperature $\beta$ for various values of the anharmonicity parameter $\lambda$ (right panel).

Figure 7. Average values of the energy of the QAO in dependence on the anharmonicity parameter $\lambda$ for various values of inverse temperature $\beta$ (left panel) and on the inverse temperature
$\beta$ for various values of the anharmonicity parameter $\lambda$ (right panel).

\end{document}